\documentclass[a4paper]{jpconf}
\usepackage{graphicx}
\begin{document}
  \title{Test of the ATLAS pion calibration scheme in the ATLAS combined test beam}
\author{Francesco Span\`o~\footnote{On behalf of the Combined Test Beam and Hadron Calibration groups of the ATLAS collaboration.} }
\address{Columbia University, Nevis Laboratories, 136 South Broadway, P. O. Box 137, Irvington, NY 10533, USA }
\ead{francesco.spano@cern.ch}
\begin{abstract}
Pion energy reconstruction is studied using the data collected during the 2004 ATLAS combined test beam.  The strategy to extract corrections for the
non-compensating nature of the ATLAS calorimeters  for dead material losses and for leakage effects is discussed and assessed. The default ATLAS strategy
based on a weighting technique of the energy deposits in calorimeter cells 
is presented and compared to a novel technique exploiting correlations among energy deposited in calorimeter layers.
\end{abstract}

\section{Introduction}

Non-linearity and resolutions degradation in energy reconstruction of
hadrons by calorimeters result from non-compensation effects compounded by unmeasured energy
deposited in non-instrumented (dead) material. Calibration techniques are used to recover 
linearity and improve resolution.

In the year 2004 the ATLAS collaboration carried out a test-beam where
a full central~\footnote{The part of the ATLAS detector that is
  closest to the proton-proton interaction point in terms of
  pseudo-rapidity (defined in footnote~\ref{foot:pseudo}).} slice of the ATLAS detector was exposed to beams of
electrons and pions a large energy range. One of the main purposes of
this combined test-beam was to test the ATLAS strategy to use a
calibration based on simulation to reconstruct the correct energy of pions.

The general long-term ATLAS strategy to use simulation to
calibrate the detector response to hadrons is shown
in~\cite{pospelov}, while the quality of the simulation for the
combined test beam data is discussed in~\cite{speckmayer}. The result of these
two inputs is shown in the present report.

The experimental setup  and the real and simulated data samples are
briefly illustrated in sections~\ref{sec:expset} and~\ref{sec:dataandsim}.
The pion calibration techniques are illustrated in
section~\ref{sec:calibtech}. The performance on simulation and 
data is discussed in terms of linearity and resolution in 
section~\ref{sec:perf}.

\section{Experimental setup}
\label{sec:expset}

\begin{figure}
\begin{center}
\includegraphics[width=24pc]{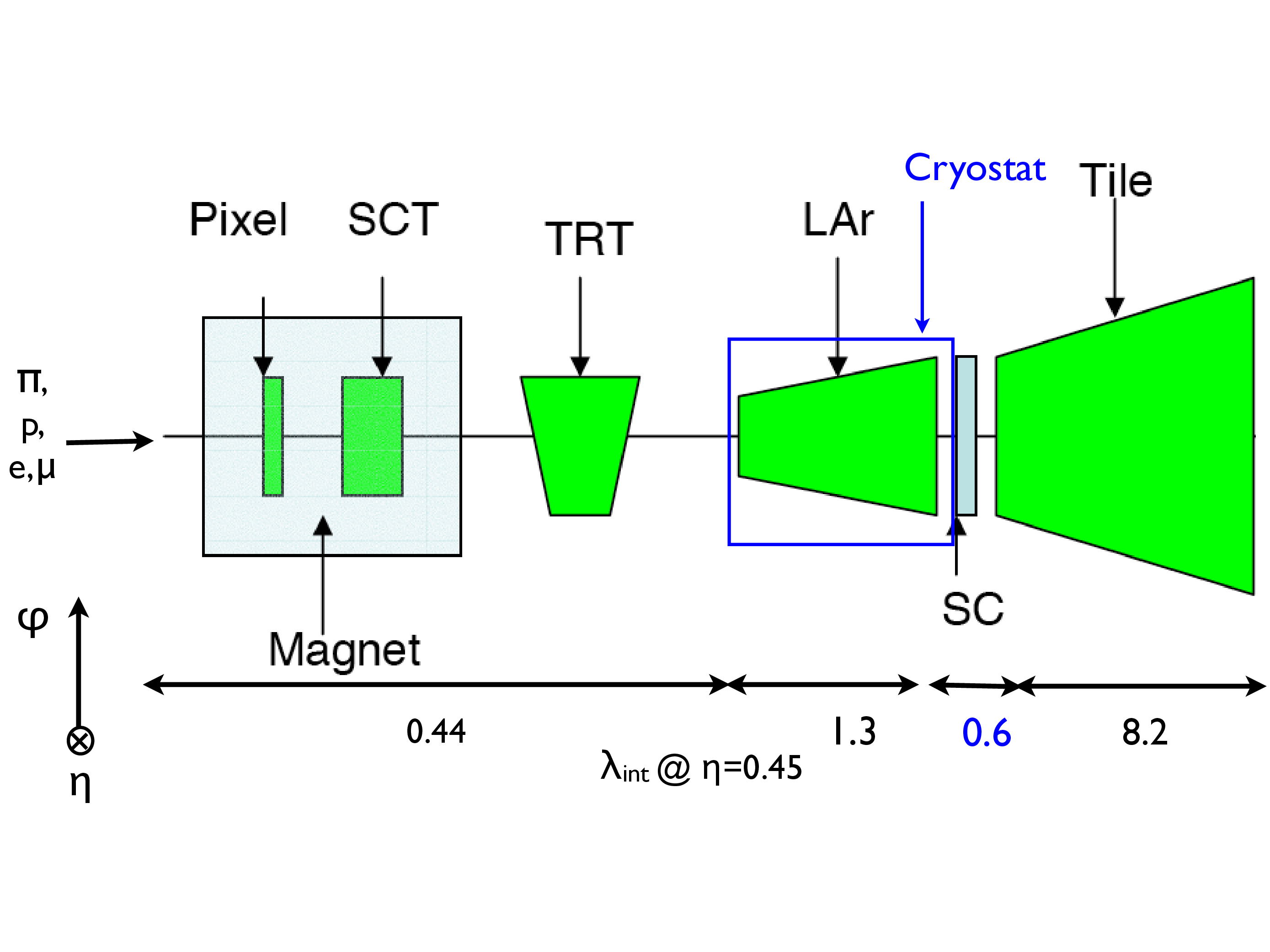}
\end{center}
\caption{\label{fig:ctb04setup} The experimental layout for the 2004 ATLAS combined test beam. See text for details.}
\end{figure}

The 2004 ATLAS combined test beam is shown in the sketch of
figure~\ref{fig:ctb04setup}. It was composed of a full central slice of 
the ATLAS detector extending for about three units in pseudo-rapidity~\footnote{Pseudo-rapidity ($\eta$) is defined at $\eta$ =  -log(tan($\theta$/2)) where $\theta$ is the polar angle in the detector, shown in fig~\ref{fig:ctb04setup}.\label{foot:pseudo}} and for 0.3 radians the azimuthal direction, $\phi$, around the beam axis.
The central semiconductor pixel and strip detectors were housed in a 
bending magnet and followed by the straw-tube transition radiation
tracker. The ATLAS central sampling calorimeters followed: one barrel
module of the liquid argon-lead  electromagnetic calorimeter (LAr) with
accordion shape was housed in its cryostat and put in front of three
hadronic iron-scintillator modules (Tile) stacked in the azimuthal direction, 
orthogonally to the incoming test beam axis.
The setup was exposed to beams of particles (pions, protons, electrons
and muons) in the energy range 1 to 350 GeV. At $\eta$ = 0.45 the
material in front of the calorimeters is estimated to consist of 0.44
$\lambda_{int}$~\footnote{The interaction length is calculated for
  protons.} and the calorimetry stretches for about 9.5
$\lambda_{int}$ : 1.3 $\lambda_{int}$ for LAr and 8.2 $\lambda_{int}$ for Tile.
The LAr cryostat accounts for additional 0.6 $\lambda_{int}$
in between the LAr and the Tile.

\section{Data and simulation samples}
\label{sec:dataandsim}
The data consists of samples of events in which positive pions impinge
on the experimental setup at $\phi$ = 0 and $\eta$ = 0.45. They are
summarized in table~\ref{table:data}.

\begin{table}
\caption{\label{table:data} Properties of positive pion data samples used in the analysis. The sample fraction of proton
  events is also reported. At 180 GeV, pion events are
 selected from a sample of positron events.}
\begin{center}
\begin{tabular}{llll}
\br
Positive pion data samples& & & \\
\mr
Selected events & Energy (GeV) & Proton contam. (\%) \\
8000 & 20 & 0 \\
15000& 50& 41 \\
7000& 100 & 59 \\
5000 & 180 & 75  \\
\br
\end{tabular}
\end{center}
\end{table}

 The pion beams are generated from proton primary beams extracted from
 the CERN SPS accelerator: the resulting proton contamination is
 measured by estimating the fraction of proton events that are necessary to
 reproduce the observed probability of generating a high energy hit in
 the transition radiation detector.
The pion selection is documented in~\cite{speckmayer}.
The proton contamination required simulation of samples of
pions and protons in the range 15 to 230 GeV with GEANT 4.7~\cite{geant} using the
QGSP\_BERT physics~\cite{qgspbertini} list and a consistent
 description of the test-beam set-up. The 4 million events simulated
 were split in two statistically independent sets of samples, the
 first one used for deriving simulation corrections and the second for testing the expected performance.

\section{Pion calibration techniques}
\label{sec:calibtech}

An incoming pion in the ATLAS detector causes a shower that is sampled
by the seven calorimeter layers of the combined electromagnetic and
hadronic sections. Any hadronic calibration scheme has to recover the 
intrinsic losses due to the invisible energy lost in nuclear
interactions. Additional imperfections in the reconstruction also need
to be accounted for: corrections are required for the imperfect energy
collection of the clustering algorithm (out-of cluster) and incomplete shower
containment (leakage of neutrons, muons and neutrinos). Finally 
corrections for energy deposits in non-instrumented material have an
important role. In particular, in the ATLAS central region (barrel)
a non-negligible amount of dead material is present between 
the electromagnetic and hadronic compartments i.e. in the midst of the 
longitudinal development of most hadronic showers.

Two calibration techniques are considered.
The default ATLAS local hadronic calibration (LH in the following) is described in detail in~\cite{pospelov}. 
A novel technique (described in section~\ref{sec:calibtech}) is also considered: it
is based on the use of correlations between the signals at the layer level, summing the
clustered energy in each calorimeter longitudinal segment. It is called layer correlation calibration (LC in the following).
The ansatz is that hadronic and electromagnetic energy deposits have
different fluctuations properties and, consequently, variables that
are sensitive to fluctuations in the total energy can be used both to
derive all the corrections  and improve the resolution of the total 
energy measurement. 

The two techniques result in different outputs: LH produces calibrated
clusters that will be used to form calibrated jets. On the other hand LC provides calibrated
layer energies: such scheme is technically extendible to jets, but 
it will be aimed at calibrating the given jet energy depositions in a
layer.

\section{Performance of the hadronic calibration schemes}
\label{sec:perf}
Performance is assessed in terms of linearity and relative
resolution. 

The total energy is fitted with a Gaussian in the
[$\mu$- 2$\sigma$, $\mu$ + 2$\sigma$] interval where $\mu$ is the mean of the
initial energy distribution and $\sigma$ is its standard deviation.
Linearity is defined as the ratio of the expected fitted average to
the beam energy as a function of the beam energy.
Relative resolution is defined as the ratio of the fitted standard
deviation to the fitted average as a function of the beam energy.

\subsection{Expected linearity and resolution for the LH scheme}
\label{sec:hcalexpperf}

\begin{figure}
\begin{center}
\includegraphics[width=24pc]{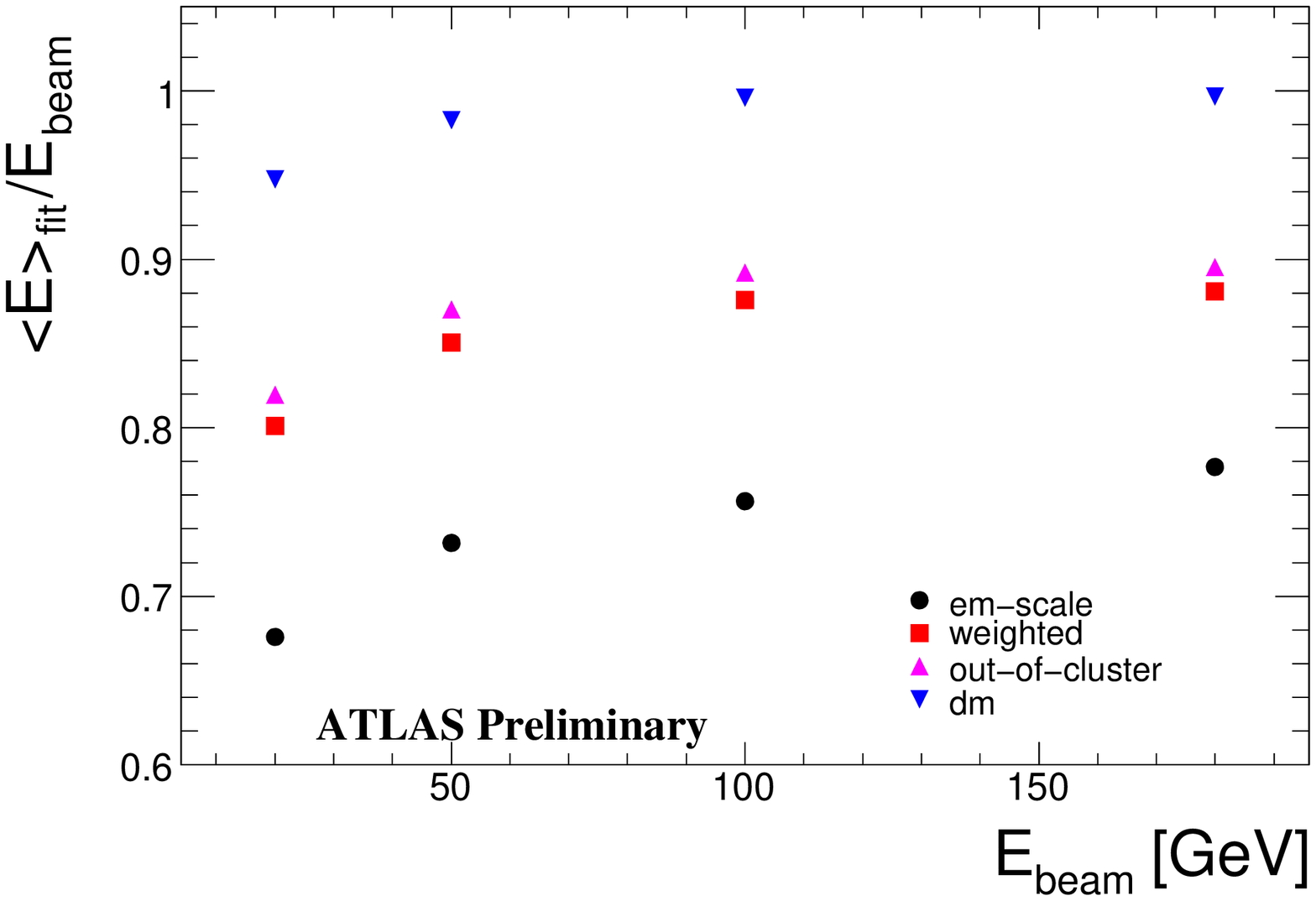} 
\includegraphics[width=24pc]{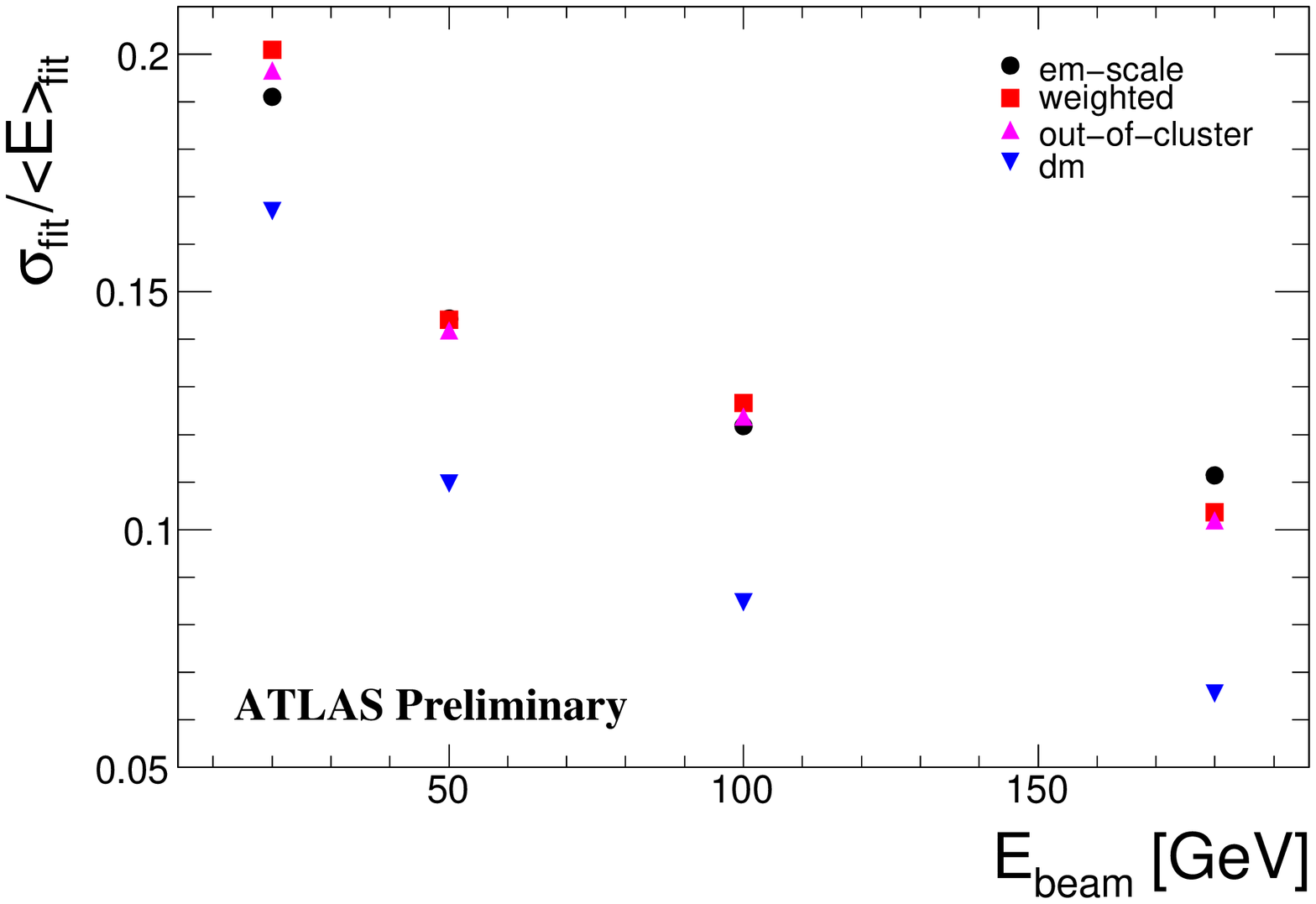} 
\end{center}
\caption{\label{fig:linhadcal}LH scheme: linearity (upper plot) and relative resolution (lower plot) for simulated positive pion
  events impinging on the test-beam setup at $\eta$ = 0.45. The various
  stages of correction are shown (see text for details).}
\end{figure}

Linearity obtained with LH from simulated positive pion events is
shown in the upper plot of figure~\ref{fig:linhadcal}.
At the electromagnetic scale the typical linearity shape for 
non-compensating calorimeters is observed: about 75\% of the beam
energy is measured and the linearity ratio increases with beam energy 
due to the increasing electromagnetic fraction of the shower.
The compensation weights recover about 10\% of the total beam energy.
The small out-of-cluster corrections account for about 1\% of the beam
energy. The remaining 10\% is recovered by adding the dead material
corrections. Linearity is finally recovered within 2\% for beam energy larger than
20 GeV. 

The relative resolution is shown in the lower plot of figure~\ref{fig:linhadcal}.
Dead material effects are expected to play a dominant role.
The improvement in relative resolution deriving from suppressing the
various fluctuations is expected to reach 11\% to 40\%.


\subsection{LC scheme}
\label{sec:lcdescr}
The LC technique defines the total pion energy as the sum of
clustered energy for each calorimeter layer.
The event-by-event layer energy corrections are defined as a function
of a specific pair of linear combinations of layer
energies. Such combinations are the components of the
seven-dimensional vector of layer energies along the vector space
basis derived by a principal component analysis (PCA)~\cite{PCA}: the two
components are used along the basis vectors whose associated PCA variance gives the largest
contributions to the fluctuations of the total energy. Intuitively,
the corrections depend on the ``directions of largest independent
fluctuations'', $E_{eig,i}$, where $i$ is an integer from one to seven. 

Both the weights aimed at compensating the invisible energy and 
the corrections aimed at recovering dead material losses are derived
as two dimensional look-up tables (seven for the layer energies, one for the
dead material correction). For both types of tables, $E_{eig,0}$, the linear
combination of layer energies with the largest contributions to the
total energy fluctuation is one of the two dimension.

\begin{figure}
\begin{center}
\includegraphics[width=24pc]{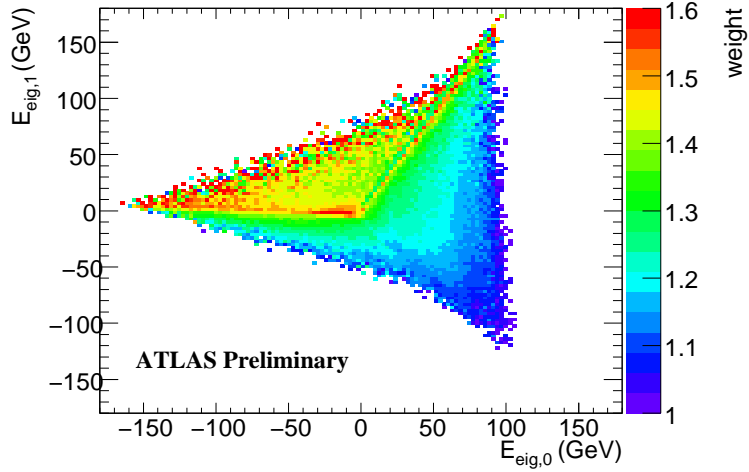}
\end{center}
\caption{\label{fig:tileweight}LC scheme: two dimensional look-up table of
  compensation weights for the first layer of the Tile hadronic
  calorimeter (see details in the text). Pions and protons samples are mixed to achieve a 41\% proton contamination.
}
\end{figure}

$E_{eig,1}$, the linear combination  with the second largest contributions to the total
energy fluctuation, is the second variable for the calculation of the
compensation weights. The weight for a given layer energy in a given
bin of the two dimensional table is defined as
the average $E_{true}^{layer}/E_{rec}^{layer}$ over all the events in the bin~\footnote{Such
  correction includes out-of-cluster effects as the numerator is the true total energy
  deposited in the layer.}. An example of the weight table for
the first Tile layer is shown in figure~\ref{fig:tileweight}.
The separation between the high weight region, dominated by invisible
energy, and the low-weight region, dominated by visible energy, is
evident.

\begin{figure}
\begin{center}
\includegraphics[width=24pc]{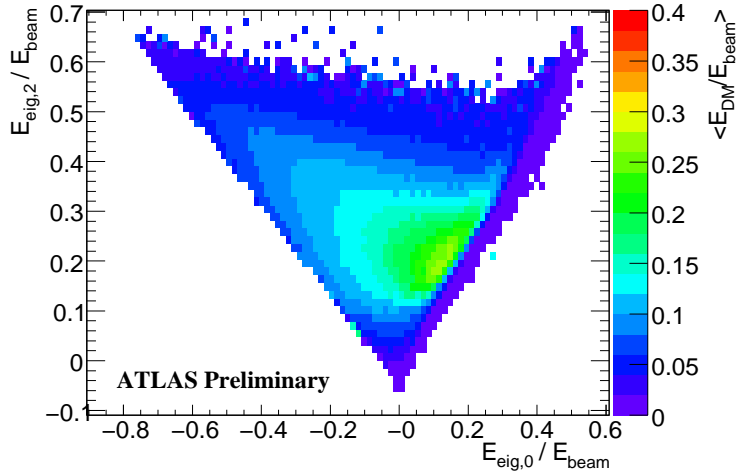}
\end{center}
\caption{\label{fig:dmtot}LC scheme: two dimensional look-up table of
  corrections for energy loss in dead material (as a fraction of total
  energy) between TileCal and LAr electromagnetic calorimeter (see
  details in the text). Pions and prootn samples are mixed to achieve
  a 41\% proton contamination.}
\end{figure}

For the dead material correction, the second dimension is represented by, $E_{eig,2}$,the
layer energy combination with the third largest contribution to the 
total energy fluctuations~\footnote{For dead material corrections this combination is found to have
  the best expected performance.}. The look-up table is shown in figure~\ref{fig:dmtot}
where the high dead material correction region is well separated form
the rest. The correction is derived as a function of the
normalized linear combinations mentioned above (each combination is divided by the best
estimate of the total energy) and it is expressed as a fraction of the
total energy itself.

A small correction for the leakage, dead material energy losses upstream of the
calorimeters and in between the first (presampler) and second
(``strips'')  LAr layers is calculated by a parametrization obtained
from simulation as a function
of the total energy estimate $E_{tot}$~\footnote{The formula
  is $E^{DM}_{func}(E) = C_{upstream}(E) E_{LArpresampler}+ C_{LArpresampler-strips}(E)\sqrt{|E_{LArpressampler}E_{LArstrips}|}+C_{leakage}(E).$
}.

An iterative procedure is then applied: a given total energy estimate provides
a new dead material correction which can in turn be used to determine
the total energy. A few iterations are required to obtain a stable result.

All look-up tables are filled by the full set of simulated samples
from 15 to 230 GeV so as to reduce the beam energy dependence of the
correction as much as possible.

\subsection{Linearity and resolution for the LC scheme}
\label{sec:lcperf}
In the case of the LC scheme a mix of pions and
protons was used to derive the corrections and to simulate the data.
The contamination values are those from table~\ref{table:data}. 

\begin{figure}[htbp]
\begin{center}
\includegraphics[width=35pc]{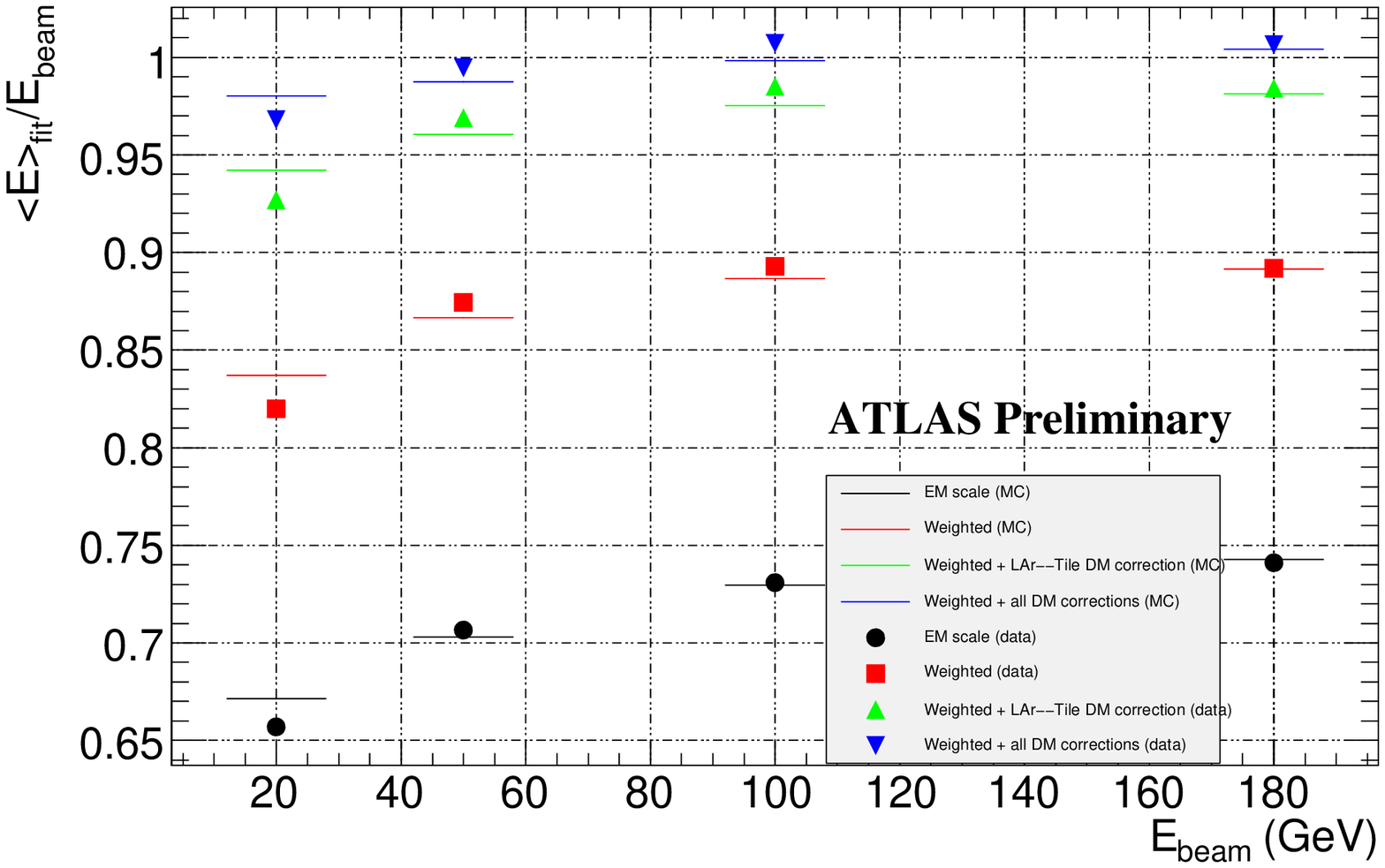}
\includegraphics[width=35pc]{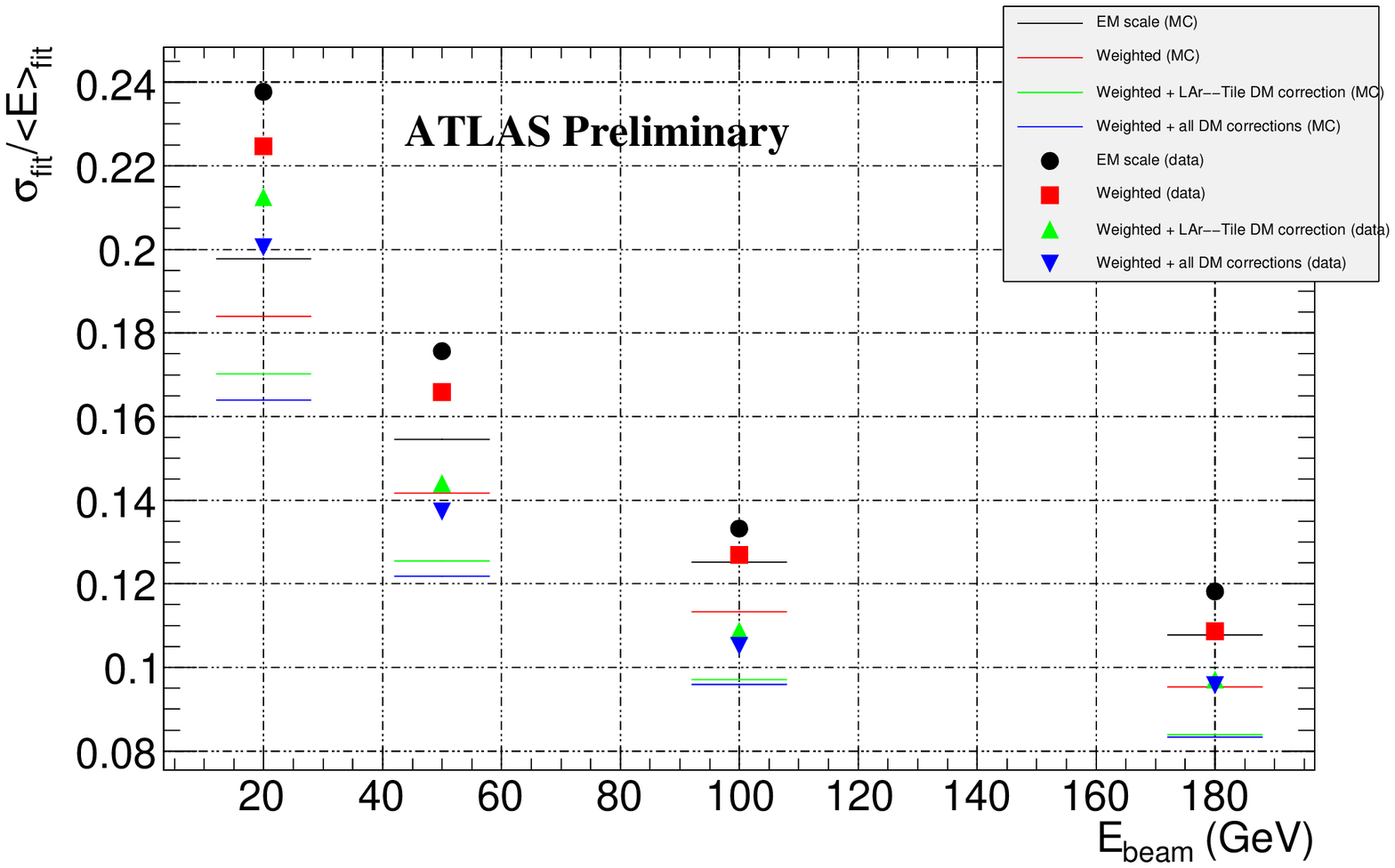}

\end{center}
\caption{\label{fig:lclinres}LC scheme: linearity (upper plot) and relative resolution (lower plot) 
   for positive pion events impinging on the test-beam setup at $\eta$ = 0.45 at the
  various stages of correction (see text for details). Data are
  shown in filled symbols, simulated events are represented by
  horizontal lines.}
\end{figure}

The upper plot of figure~\ref{fig:lclinres} shows the linearity obtained for both data and
simulation: the agreement is within 2\% at all stages of calibration.
The resulting picture is similar to that outlined in section~\ref{sec:hcalexpperf}
for the LH scheme.
The reconstruction at the electromagnetic scale is  accounting
for 75\% of the beam energy. The compensation weights recover about 12\% of the 
beam energy while the dead material correction accounts for about 10\%.
The dead material correction for losses between Tile and LAr 
represents about 80\% of the total dead material
corrections.
The LC method recovers linearity within 3\% over the whole energy
range.

The relative resolution is shown the lower plot of figure ~\ref{fig:lclinres}.
The simulation foresees a relative improvement of 17 to 24\%: the
data behave consistently showing an improvement of 17 to 21\%.
Even though the relative behaviour is the same, the simulation
underestimates the resolution in the data by about 25\%.
GEANT4.9~\cite{speckmayer} is expected to improve the data description.

%

\subsection{LH vs LC scheme: linearity}

\begin{figure}
\begin{center}
\includegraphics[width=36pc]{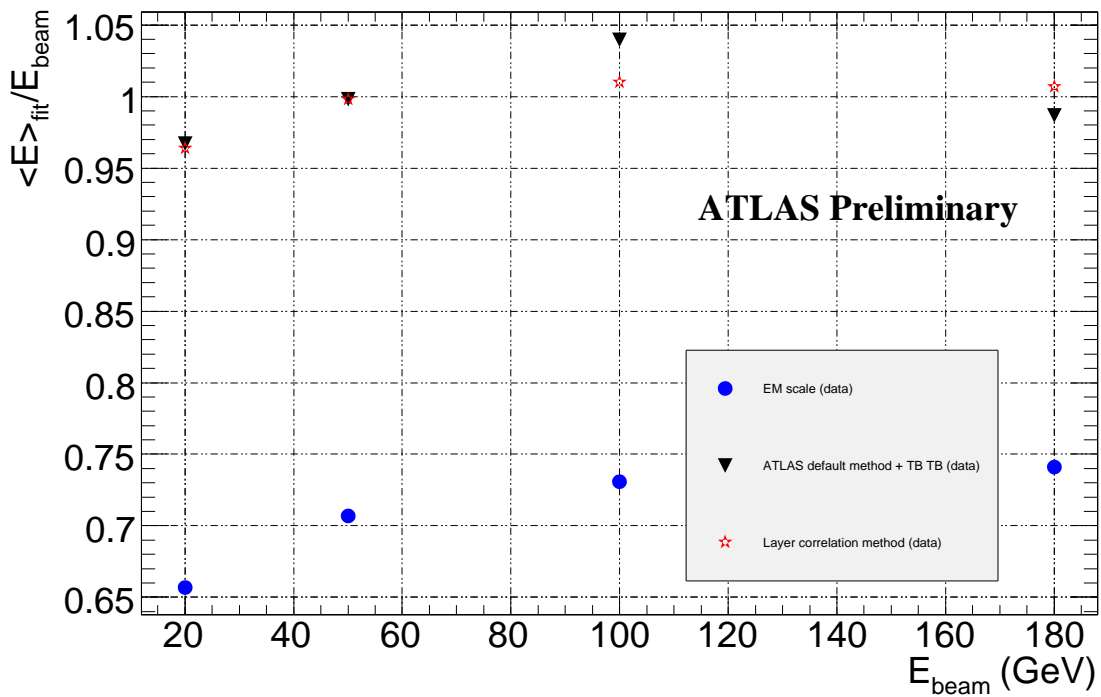}
\end{center}
\caption{\label{fig:lcvslhlin} Comparison of linearity for the LH and
  LC schemes. Only the electromagnetic scale and the fully corrected
  scale are shown.}
\end{figure}

Figure~\ref{fig:lcvslhlin} shows the comparison for the resulting 
linearity when applying both calibration techniques to the same data set~\footnote{The
  electromagnetic scale result is taken from the LC
  method: a slightly different event selection was applied to the data
  when applying the LH scheme.}
 The result is quite consistent: the linearity is recovered  within 2 to 5\% by both techniques.
 
\newpage

\section{Conclusions}


%
%
%
%
A simulation-based cell-weighting technique for hadronic signal calibration was
applied to pion energy reconstruction in 2004 ATLAS combined test beam
for beam energy in the range 20 to 180 GeV. 
A novel technique based on the correlation amongst layer energies was also used.

 The linearity of response to charged pions is recovered within 2
to 5\% by both approaches in good agreement between data and
simulation; compensation weights and dead material effects have similar
impact. According to simulation, the relative energy resolution is expected to improve (by 20-30\%
to 40\%). LC actually achieves an improvement of 17 to 21\%. Simulation underestimates data resolution by 10 to 25\%;
dead material effects are dominant.

 Data-simulation discrepancies at the electromagnetic scale keep their
 size at all stages of calibration, thus simulation performance is the
 limiting factor.

\ack

The essential collaboration of Tancredi Carli, Karl-Johan Grahn and
Peter Speckmayer is gratefully acknowledged.


\end{document}